\documentclass[pra,twocolumn,10pt,aps,nofootinbib,superscriptaddress,showkeys,showpacs]{revtex4-1}
\usepackage{amsmath,amsfonts,amssymb,amstext,amscd,amsthm,dsfont,lipsum}
\usepackage{physics}
\usepackage{color}
\usepackage{soul,xcolor}
\usepackage{gensymb}
\usepackage[normalem]{ulem}
\usepackage{lipsum}

\usepackage[T1]{fontenc}

\usepackage{graphicx}

\usepackage{braket}
\usepackage{placeins}

\DeclareMathOperator{\sinc}{sinc}

\newcommand{\mypm}

\begin{document}

\title{Influence of coincidence detection through free-space atmospheric turbulence using partial spatial coherence}

\author{Samkelisiwe Purity Phehlukwayo} 
\affiliation{Quantum Research Group, School  of Chemistry and Physics,
  University of KwaZulu-Natal, Durban 4001, South Africa}

\author{Marie Louise Umuhire} 
\affiliation{Quantum Research Group, School  of Chemistry and Physics,
  University of KwaZulu-Natal, Durban 4001, South Africa}

\author{Yaseera Ismail} 
\email{Ismaily@ukzn.ac.za}
\affiliation{Quantum Research Group, School  of Chemistry and Physics,
  University of KwaZulu-Natal, Durban 4001, South Africa}

\author{Stuti Joshi} 
\email{stuti13@physics.iitd.ac.in}
\affiliation{Department of Physics, Indian Institute of Technology Delhi, Hauz Khas, New Delhi 110016 India}

\author{Francesco Petruccione}
\email{petruccione@ukzn.ac.za}
\affiliation{Quantum Research Group, School  of Chemistry and Physics,
  University of KwaZulu-Natal, Durban 4001, South Africa}
\affiliation{ National
  Institute  for Theoretical  Physics  (NITheP), KwaZulu-Natal,  South
  Africa}


\begin{abstract}
The development of a quantum network relies on the advances of hybrid systems which includes ground to ground communication.  However, the atmospheric turbulence of the environment poses a severe challenge to the optical quantum link. In this paper, we outline a theoretical and experimental investigation of the influence of atmospheric turbulence on the coincidence detection of the entangled photon pairs using a fully and partially spatially coherent pump beam. A spatial light modulator is encoded with Kolmogrov model to mimic the atmospheric turbulence strength. The results show that the photon pairs generated using a partially spatially coherent pump are more robust towards varying atmospheric turbulence strengths than the photon pairs produced by a fully spatially coherent pump beam.  
\end{abstract}

\maketitle

\section{ Introduction}

Quantum communication is a promising physical process of upholding the security of information. It is considered the most secure method of data transfer, based on the encoding of single photons that are transmitted across  fibre or a free-space link, which includes ground and satellite-based communication \cite{ziellinger2007, pan2012, Yin2017, pan2017, gisin2010, guo2008}. At the core of the development of quantum technology is the phenomena of entanglement. In photons, for example, entanglement may be encoded within all accessible degrees of freedom, such as the polarization, spatial and spectral properties \cite{gisin1999,kwait1995,boyd2004}. The generation (and distribution) of entanglement between photons  generally forms the basis of modern quantum communication protocols.

Free-space quantum communication, in the spatial regime, is severely hindered by the influence of atmospheric turbulence. Turbulence is known to destroy the entanglement with spatial modes; here are summarized the problems and findings \cite{semenov2010entanglement,jha2010effects,avetisyan2016higher,zhang2016experimentally}. In the case of spatial modes, the decay of entanglement for the evolution of a qubit (quantum bit) pair in turbulence has been studied both theoretically and experimentally \cite{semenov2010entanglement,jha2010effects,avetisyan2016higher,zhang2016experimentally}. Transmitting entangled photons, generated through spatial modes, through a free-space quantum channel results in the spatial modes being affected by the atmospheric turbulence, which reduces the probability of detection. Furthermore, the imposed scattering amongst spatial modes leads to a loss of entanglement in the final state measured in a specified subspace. Recently, it has also been shown that the turbulence also influences the detection scheme of polarization based entanglement \cite{ismail2017}. 
Besides quantum information science, entanglement has also found applications in numerous research areas such as ghost imaging \cite{sergienko1995}, sub-wavelength interference, and microscopy \cite{d2001}. 
All the studies mentioned above have been carried out by considering the pump to be fully spatially coherent. In some applications, partially spatially coherent beams are found to portray advantages over fully spatially coherent beams. Theoretical studies have shown the transfer of the angular profile and spatial coherence of the pump into the twin-photon fields \cite{jha2010spatial}.  The experimental demonstrations of these studies confirm the effect of pump spatial coherence on the polarization-entangled \cite{ismail2017polarization} spatial coherence and entanglement properties of the down-converted field \cite{HugoDefienne2019}. Entangled photons generated using a partially coherent pump are more robust to  atmospheric turbulence than the entangled photons generated using a fully spatially coherent pump \cite{qiu2012}. Generation of entangled photons using a partially coherent pump beam provides a flexible way to control entanglement between twin photons by the pump parameters.
\newline
\indent
In the present work, a theoretical and experimental study is carried out to investigate the influence of atmospheric turbulence on the coincidence between signal and idler photon pairs. Our study is based on using a spatial light modulator  encoded with a Kolmogorov model of turbulance \cite{kolmogorov21941,kolmogorov1941} to mimic such effects. It is observed that the coincidence counts decreases as the atmospheric turbulence strength increases for different propagation distances. The observations show the coincidence counts between the signal-idler photons generated using partially spatially coherent pump beam (PSCPB) are less affected by the atmospheric turbulence than the coincidence counts between signal-idler photons generated using fully spatially coherent pump beam (FSCPB). To the best of our knowledge, this is the first experimental verification of a partially spatially coherent pump source of entangled photons propagating through atmospheric turbulence. Present work shows the robustness of down-converted photons, generated using PSCPB, towards varying atmospheric turbulence strength which would be useful in the practical design of entangled-photon based free-space quantum communication systems and quantum key distribution.

\section{Theoretical background}

\begin{figure}
	\centering
	\includegraphics[width=\linewidth]{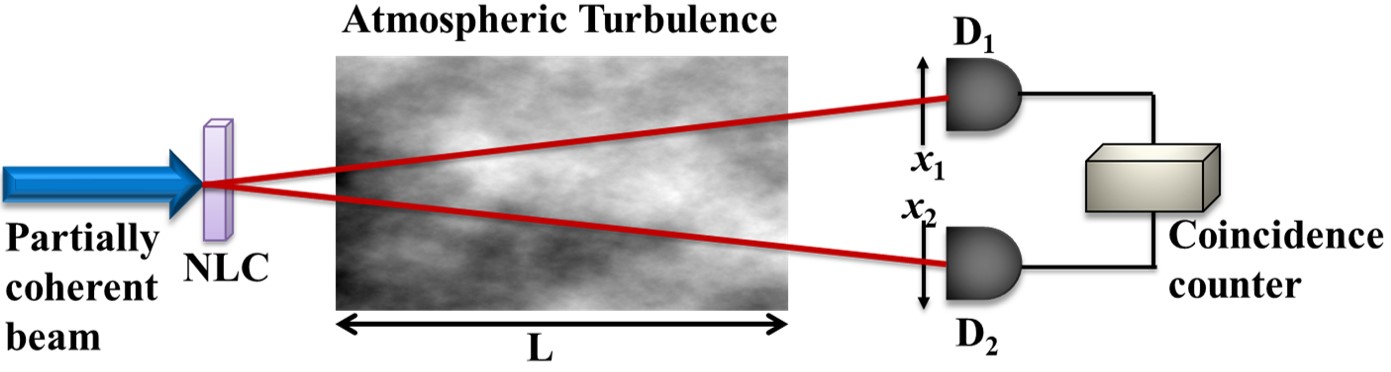}
	\caption{Schematic to study the effect of atmospheric turbulence on entangled photons generated using partially spatially coherent pump beam.}
	\label{fig:scheme}
\end{figure}

Free-space quantum communication entails the transfer of the generated single photons across a quantum channel which can vary in atmospheric turbulence and propagation distance. The generic scenario to study the influences of atmospheric turbulence on the entangled-photons generated using a  PSCPB is shown in Fig. \ref{fig:scheme}. The partially spatially coherent (PSC) pump interacts with a nonlinear crystal, responsible for Spontaneous Parametric Down Conversion (SPDC), and generates entangled photons according to the laws of conservation of energy and momentum \cite{burnham1970}. This implies that photons from a pump beam produce a single photon pair which is a signal and an idler.  The generated photons reach the detectors $D_1$ (positioned at $x_1$) and $D_2$ (positioned at $x_2$) after propagating a distance $z$ through atmospheric turbulence, which represents the quantum link. Throughout the manuscript $p$, $s$ and $i$ refer to pump, signal and idler respectively. The two-photon wavefunction is given by \cite{saleh2000duality},
\begin{equation} 
\psi(x_1,x_2)=\int\int{dq_s}{dq_i}V_p(q_p)\eta(q_s,q_i)H_s(x_1,q_s)H_i(x_2,q_i),
\end{equation} 
where $V_p(q_p)$ represents the pump field and $H_s(x_1,q_s)$ ($H_i(x_2,q_i)$) is the transfer function of the signal (idler) from the crystal plane to the detector plane. The function $\eta(q_s,q_i)$ defines the phase-matching function given by \cite{saleh2000duality}, 
\begin{equation}
\eta(q_s,q_i)={\sinc}(\frac{{\Delta}q{\thinspace}L}{2})\exp(-\frac{i{\Delta}q{\thinspace}L}{2}),
\end{equation}
with,
\begin{equation}
{\Delta}q=\frac{q^2_p}{2k_p}-\frac{q^2_s}{2k_s}-\frac{q^2_i}{2k_i
},\nonumber\\
\end{equation}
where, $L$ is the length of nonlinear crystal. $k_p$, $k_s$ and $k_i$ refers the wave vector of pump, signal and idler respectively.

The coincidence count rate between the signal and the idler can be calculated from
\begin{equation} 
R(x_1,x_2)=|{\psi(x_1,x_2)}|^2.
\end{equation}
By substituting Eq.(1) into Eq.(3), and considering the pump to be PSC, the coincidence count rate can be written as
\begin{equation}\label{eq1}
\begin{split}
&R(x_1,x_2)=\int\int{dx'}{dx}{\langle}V^*_p(x')V_p(x){\rangle}\\
 &\times{\langle}h^*_s(x_1,x')h^*_i(x_2,x')h_s(x_1,x)h_i(x_2,x){\rangle},
\end{split}
\end{equation}
or
\begin{eqnarray}
R(x_1,x_2)=\int\int\int\int{d\rho_s}{d\rho_i}{d\rho'_s}{d\rho'_i}W(\rho_s,\rho_i,\rho'_s,\rho'_i)\nonumber\\
{\times}h_s(x_1,\rho_s)h_i(x_2,\rho_i)h^*_s(x_1,\rho'_s)h^*_i(x_2,\rho'_i),
\end{eqnarray}
where $W(\rho_s,\rho_i,\rho'_s,\rho'_i)$ defines the cross-correlation function of the pump which is given by 
\begin{eqnarray}
W(\rho_s,\rho_i,\rho'_s,\rho'_i)=\int\int{dx}{dx_0}{\langle}V_p(x)V^*_p(x_0){\rangle}\nonumber\\ 
{\Lambda}(\rho_s+\rho_i){\Lambda^*}(\rho'_s+\rho'_i).
\end{eqnarray}
where

\begin{equation}
x=\frac{\rho'_s+\rho'_i}{2}{\thinspace}\text{and}{\thinspace}x_0=\frac{\rho_s+\rho_i}{2}\nonumber\\
\end{equation}

While $h_j(x_1,x)$ defines the free-space impulse function \cite{goodman2005}
\begin{eqnarray}
h_j(x_1,x)=\sqrt{-\frac{ik_j}{2z\pi}}\exp(-\frac{ik_j}{2z}(x_1^2+x^2-2x_1x))\nonumber\\  
\exp(\phi_j(x_1,x)),  (j=s,i).
\end{eqnarray}

To study the influence of atmospheric turbulence experimentally, we have utilised a spatial light modulator encoded with the Kolmogorov model of turbulence. The atmospheric structure constant $C^2_n$ describes the strength of atmospheric turbulence \cite{kolmogorov21941,kolmogorov1941}. The lateral coherence length under the influence of turbulence is related to $C^2_n$ as $\alpha_j=(0.55C^2_nk^2_jz)^{-3/5}$, $j=s,i$ \cite{wang1979optical} and $k_s$ and $k_i$ is the wavevector of signal and idler respectively.

The atmospheric turbulence function of the signal and the idler is expressed as \cite{wang1979optical}
\begin{equation} \label{eq2}
\begin{split}
&\langle\exp(\phi^*_j(x_1,x')\phi_j(x_2,x))\rangle=\\
 &\exp(-\frac{(x_1-x_2)^2+(x_1-x_2)+(x'-x)^2)}{\alpha_j^2}).
\end{split}
\end{equation}
We  considered the pump to be a Gaussian Schell-model such that the spatial correlation between two transverse points $x$ and $x'$ in the pump is given by \cite{mandel1995}
\begin{equation}
{\langle}V^*_p(x')V_p(x){\rangle}=A_p\exp(-\frac{x^2+x'^2}{4\sigma^2})\exp(-\frac{(x'-x)^2}{2\delta^2}),
\end{equation}
where $\sigma$ and $\delta$ represents the beam waist and spatial coherence length of the pump beam at the crystal plane.

\begin{figure*}
	\centering
	\includegraphics[width=0.8\textwidth]{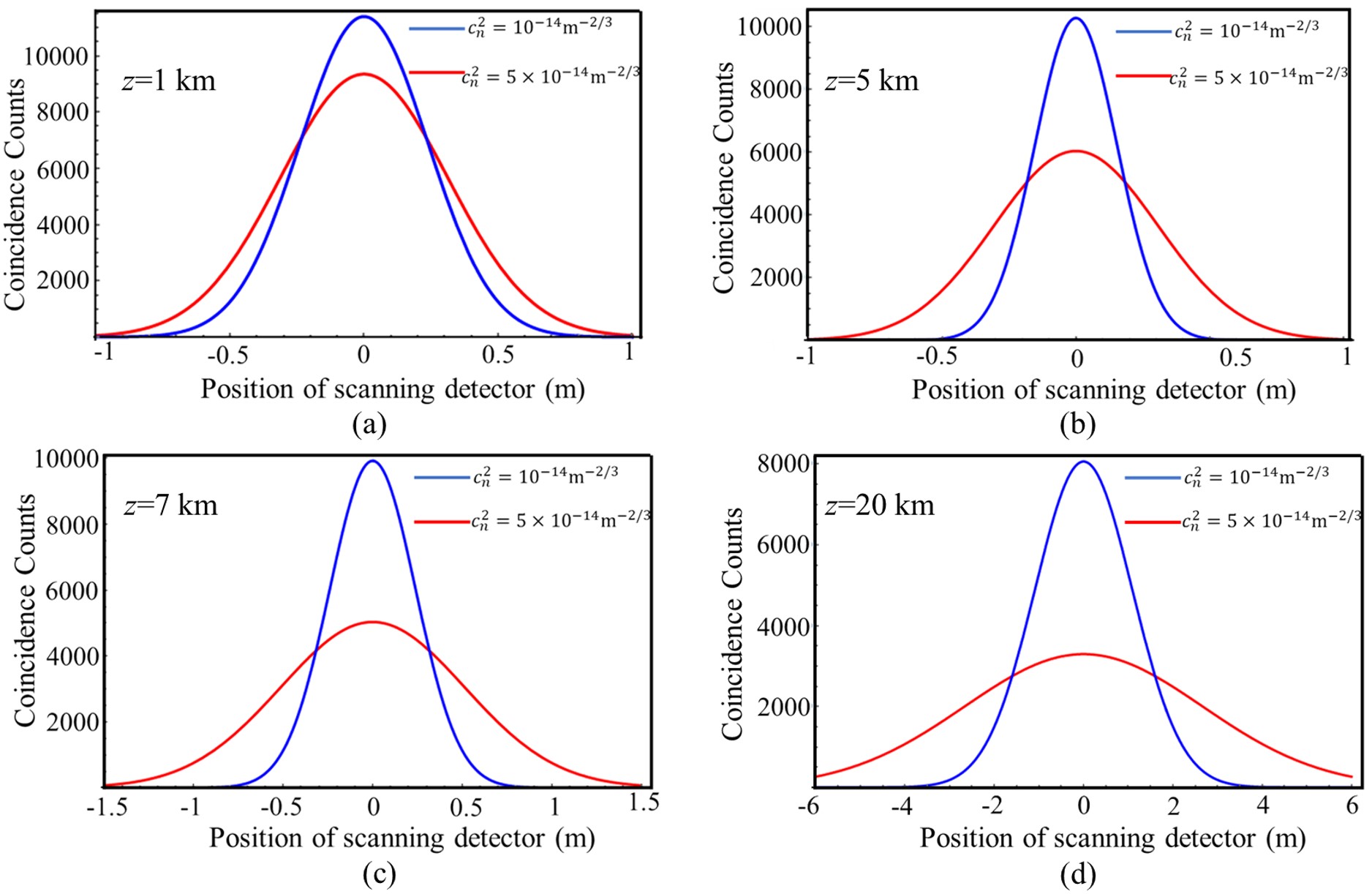}
	\caption{Theoretical plots of coincidence counts as a function of position of scanning detector for different values of atmospheric turbulence strength when pump is fully coherent $(\delta=\infty)$. (a) z=1 km, (b) z=5 km, (c) z=7 km and (d) z=20 km.}
	\label{fig:theoryfully}
\end{figure*}

\begin{figure*}
	\centering
	\includegraphics[width=0.8\textwidth]{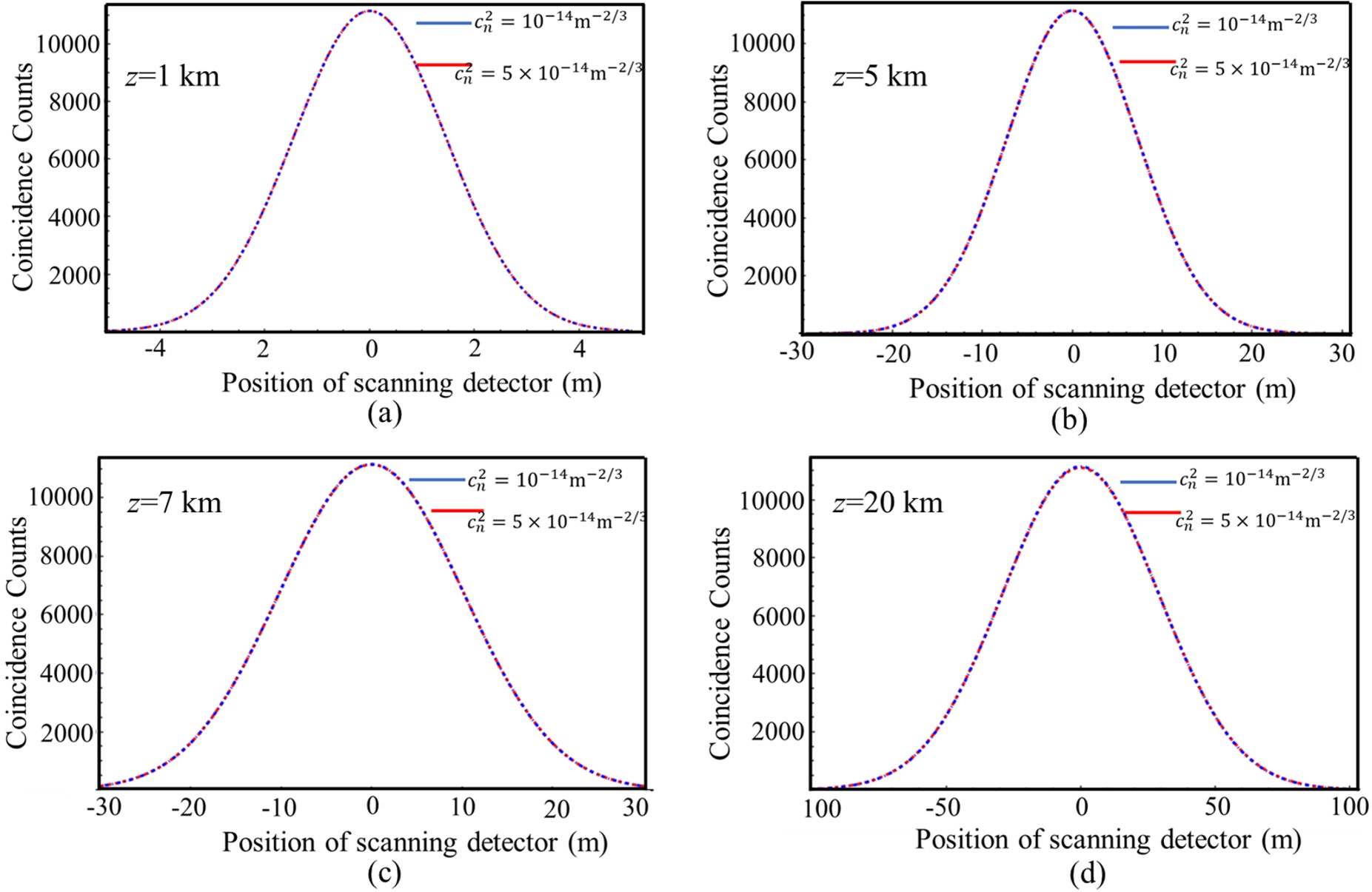}
	\caption{Theoretical plots of coincidence counts as a function of position of scanning detector for different values of atmospheric turbulence strength when pump is partially spatially coherent $(\delta=0.0876${\thinspace}mm). (a) z=1 km, (b) z=5 km, (c) z=7 km and (d) z=20 km.}
	\label{fig:theorypartial}
\end{figure*}

\begin{figure*}
	\centering
	\includegraphics[width=0.8\textwidth]{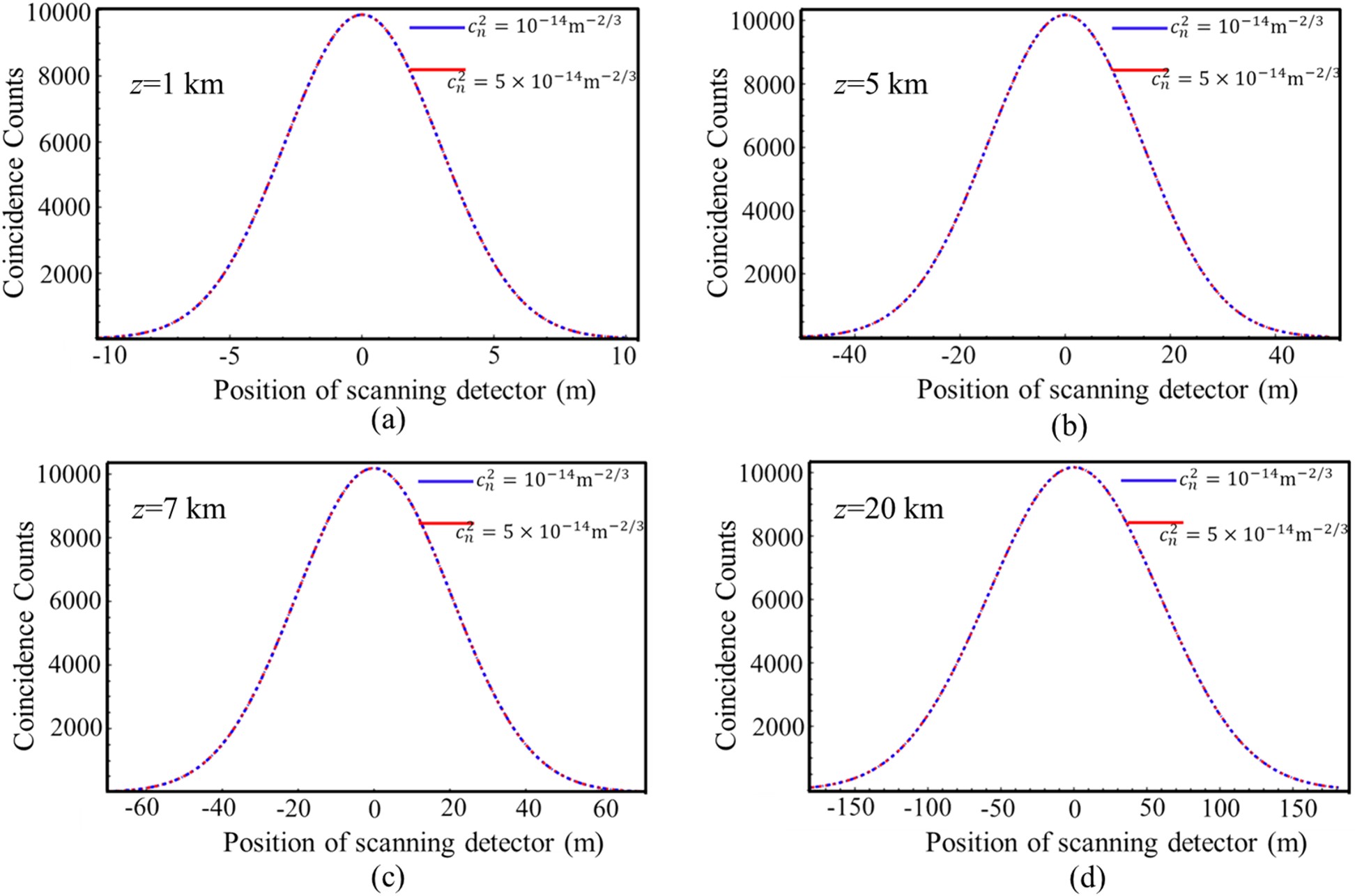}
	\caption{Theoretical plots of coincidence counts as a function of position of scanning detector for different values of atmospheric turbulence strength when pump is partially spatially coherent $(\delta=0.0417${\thinspace}mm). (a) z=1 km, (b) z=5 km, (c) z=7 km and (d) z=20 km.}
	\label{fig:theorypartial2}
\end{figure*}

\begin{figure*}
	\centering
	\includegraphics[width=0.8\textwidth]{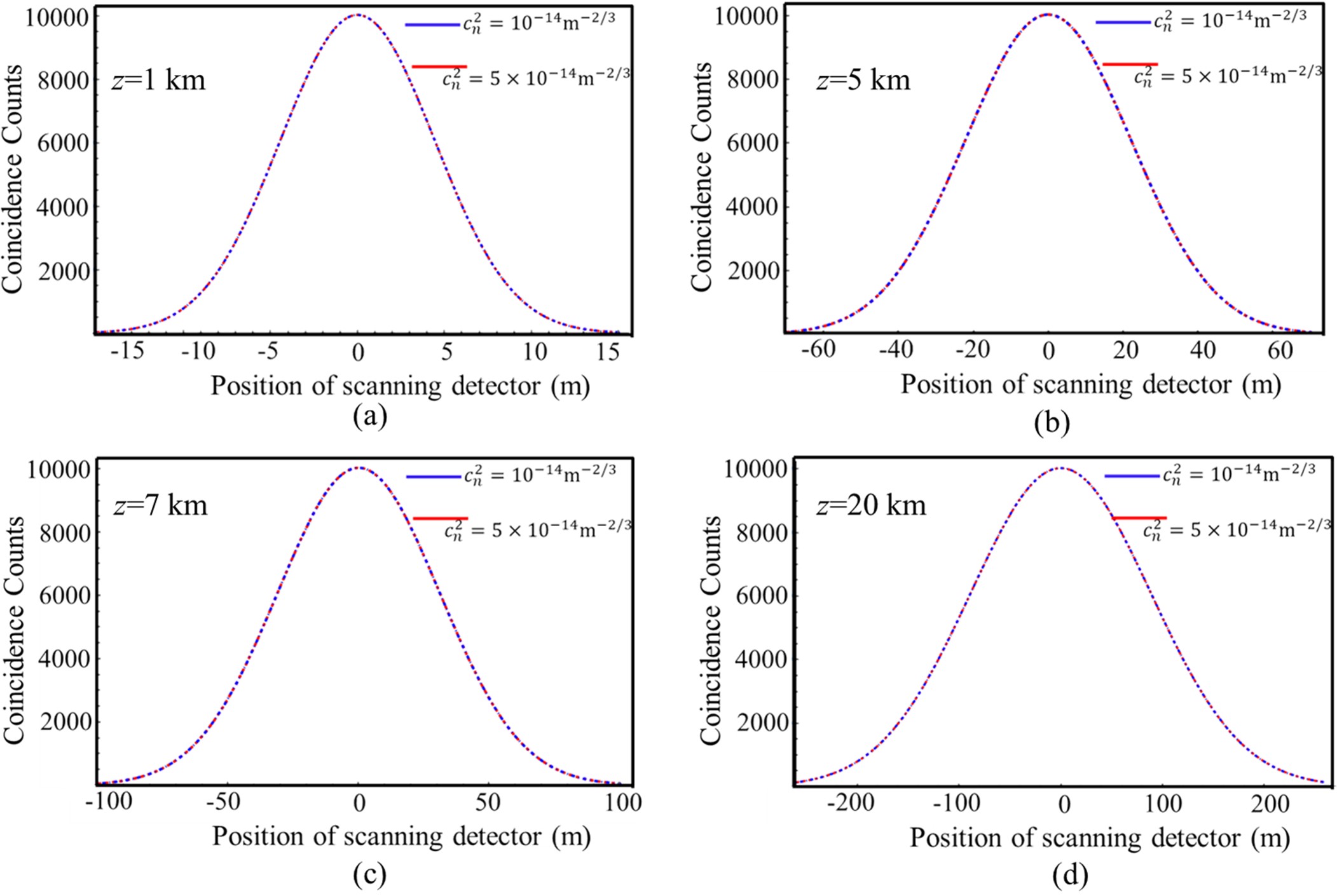}
	\caption{Theoretical plots of coincidence counts as a function of position of scanning detector for different values of atmospheric turbulence strength when pump is partially spatially coherent $(\delta=0.0253${\thinspace}mm). (a) z=1 km, (b) z=5 km, (c) z=7 km and (d) z=20 km.}
	\label{fig:theorypartial3}
\end{figure*}

\begin{widetext}

Now, by substituting Eq. (6) - Eq. (9) into Eq. (5) the coincidence count rate $R(x_1,x_2)$ can be written as 
\\


\begin{align}
 \nonumber
R(x_1,x_2) &=\frac{4{\pi}k_p}{L(\gamma^2+1)}(\frac{k_p}{4{\pi}z})^2\int\int\int\int{d\rho_s}{d\rho_i}{d\rho'_s}{d\rho'_i}\exp(-\frac{((\rho_s+\rho_i)^2+(\rho'_s+\rho'_i)^2)}{16\sigma^2})
\exp(-\frac{(\rho'_s-\rho'_i-\rho_s,\rho_i)}{8\delta^2})\\
&\times\exp(\frac{ik}{4z}(\rho^2_s-\rho'^2_s-2x_1\rho_s+2x_1\rho'_s))
\times\exp(\frac{ik}{4z}(\rho^2_i-\rho'^2_i-2x_2\rho_s+2x_2\rho'^2_i))
\times\exp(-\frac{(\rho_s-\rho'_s)^2-(\rho_i-\rho'_i)^2}{\alpha^2}).
\end{align}

Where performing each of the integrals over different parameters leads to the result


\begin{align}
 \nonumber
R(x_1,x_2) &=A\exp(-\frac{k_p^2}{16M_1z^2}(x^2_1+\frac{A^2_1x^2_1}{M_1M_3}+\frac{x^2_2M_1}{M_3}+\frac{2A_1x_1x_2}{M_3})+\frac{k^2_p}{16M_4z^2}(-1-\frac{A^2_3x^2_1}{4M^2_1}+\frac{A_3x^2_1}{M_1}+\frac{A_4x_1x_2}{M_3}-\frac{A^2_4x^2_2}{4M^2_3})) \\  \nonumber
&\times\exp(\frac{A_4k^2_px_1}{16z^2M_1M_3M_4}(A_1x_1-\frac{A_3x_2}{2}-\frac{A^2_1x_1}{4M_1M_3}-\frac{A_1A_3x_1}{2M_1}-\frac{A_1x_2}{2M_3})+\frac{1}{4M_5}(-\frac{ik_px_2}{2z}+\frac{ik_px_1}{16\delta^2M_1z}+\frac{iA_5k}{4M_3z} (\frac{A_1x_1}{M_1}))^2)\\ 
&\times\exp(\frac{iA_6k_p}{4M_1z}(\frac{A_1A_4x_1}{2M_1M_3}+\frac{A_4x_2}{2M3}-x_1+\frac{A_3x_1}{2M_1})).
\end{align}


\end{widetext}

with,

\begin{equation}
\begin{split}
&A=\frac{4{\pi}k_p}{L\sqrt{\gamma^2+1}}(\frac{k}{4{\pi}z})^2,\\
&A_1=-\frac{k_p}{4L(i+\gamma)}+\frac{1}{8\delta^2}+\frac{1}{16\sigma^2},\\
&A_2=-\frac{k_p}{4L(-i+\gamma)}+\frac{1}{8\delta^2}+\frac{1}{16\sigma^2},\\
&A_3=\frac{1}{4\delta^2}+\frac{2}{\alpha^2},{\thinspace},\\&A_4=\frac{A_1A_3}{M_1}+\frac{1}{4\delta^2},\\
&M_1=-\frac{ik}{4z}+\frac{k}{4L(-i+\gamma)}+\frac{A_3}{2}+\frac{1}{16\sigma^2},\\
&M_2=\frac{ik}{4z}+\frac{k}{4L(i+\gamma)}+\frac{A_3}{2}+\frac{1}{16\sigma^2},\\
&A_5=A_3+\frac{A_1}{4\delta^2M_1},{\thinspace}M_3=M_1-\frac{A^2_1}{M_1},\\
&M_4=M_2-\frac{A^2_3}{4M_1}-\frac{A^2_4}{4M_3},\\
&A_6=-2A_2+\frac{A_4A_5}{2M_3}+\frac{A_3\alpha}{2M_1},\\
&M_5=M_2-\frac{\alpha^2}{4M_1}-\frac{A^2_6}{4M_4}-\frac{A^2_5}{4M_3},\nonumber\\
\end{split}
\end{equation}

\begin{figure}
\includegraphics[width=0.35\textwidth]{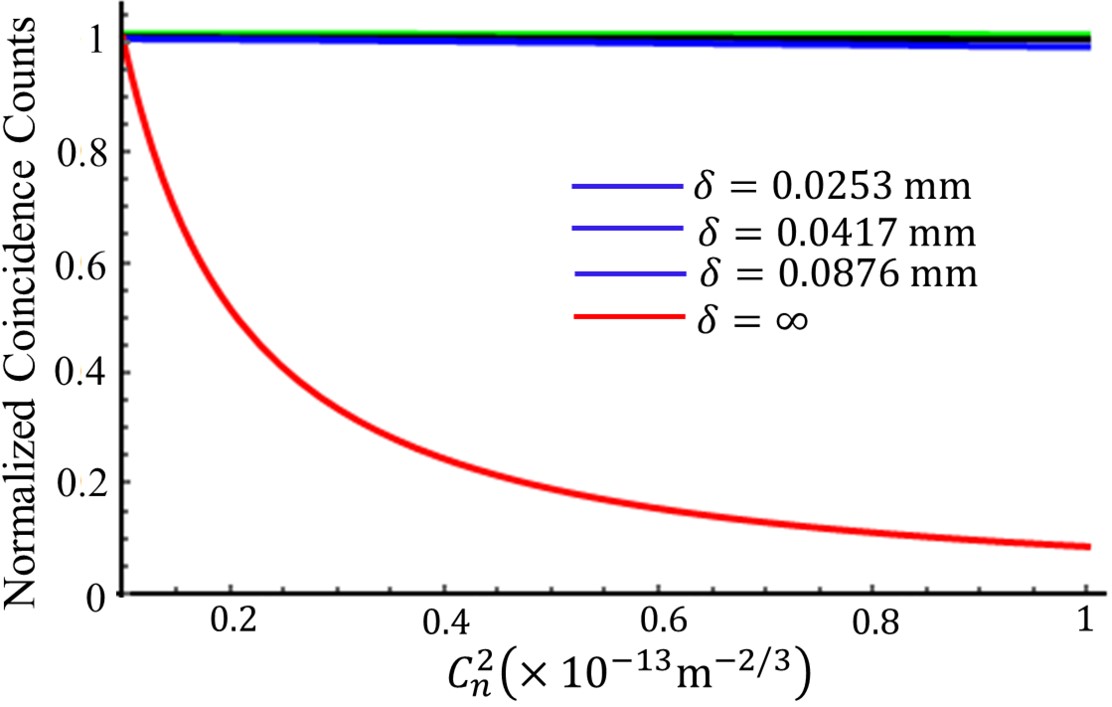}
\caption{Theoretical plots of normalized coincidence counts (at $x=y=0$) as a function of $C_n^2$ at $z=20{\thinspace}$km.}
\label{fig:varyingturb}
\end{figure}
Eq. (11) describes the coincidence count rate between the signal-idler photon pairs. The coincidence count rate depends on the pump beam parameters ($\sigma$ and $\delta$) and the strength of atmospheric turbulence for different propagation distances. Eq. (11) is used to study the influence of atmospheric turbulence on the signal-idler photon pairs. Fig. \ref{fig:theoryfully} and \ref{fig:theorypartial}  shows the atmospheric turbulence effect on the coincidence counts between signal-idler photons by considering pump to fully ($\delta=\infty$) and partially ($\delta=0.0876${\thinspace}mm) spatially coherent respectively for different propagation distances.
To mimic the propagation of single photons through a free-space optical quantum link, we emphasise the importance of the effects of varying atmospheric turbulence on entangled photons generated using the fully and partially coherent pump source. To achieve this, we have theoretically studied the detection of coincidence counts (normalised) with respect to varying atmospheric turbulence strength ($C_n^2$), which is plotted in Fig.  \ref{fig:varyingturb} using Eq. (11). The free-space propagation distance is chosen to be $z=20{\thinspace}$km. Fig.  \ref{fig:varyingturb} illustrates that at $z=20{\thinspace}$km the coincidence counts remains constant with the variation of ($C_n^2$) for a partially coherent pump source. However, there is a sharp exponential decrease in coincidence counts with ($C_n^2$) for a fully coherent pump source. In the present research, the influence of atmospheric turbulence on the coincidence counts for a fully and partially spatially coherent pump source is explored thoroughly and is followed by the experimental verification. This is the first experimental verification of a partially spatially coherent pump source of entangled photons propagating through atmospheric turbulence.

\begin{figure*}
\includegraphics[width=\textwidth]{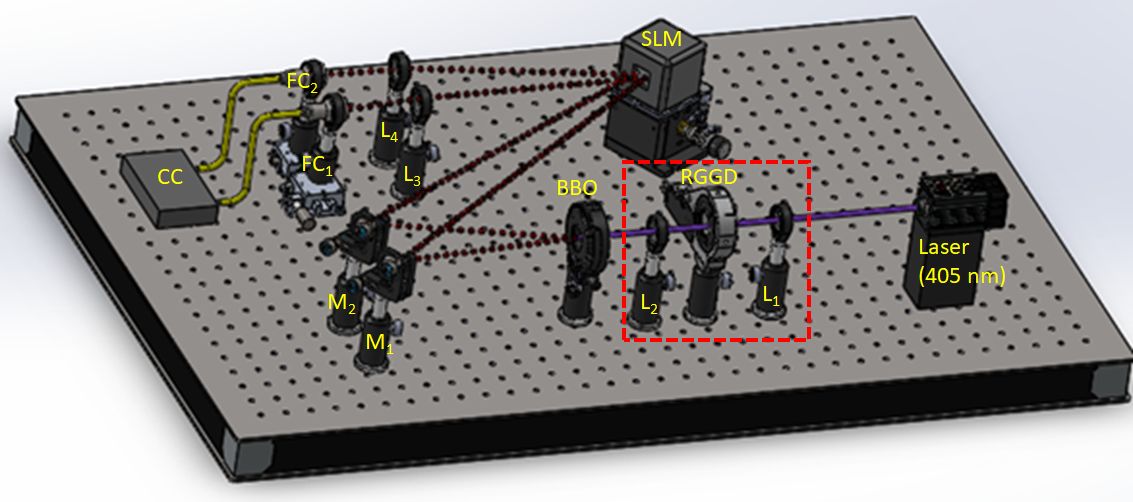}
\caption{Optical setup to verify the detection of coincidence counts for modulating partial spatial coherence in varying atmospheric turbulence. A fully coherent laser lasing at 405 nm was passed through a rotating ground glass diffuser (RGGD) and lens system ($L_1$ with a focal length 50 mm and $L_2$ with a focal length 40 mm) to produce the partial spatial coherence (dotted region). This field was passed through the type I non-linear crystal (BBO) where entangled single photon pairs at 810 nm were generated through the process of spontaneous parametric down conversion. Mirrors, $M_1$ and $M_2$, directed the photon pair towards a Holoeye $1920 {\times} 1080$ spatial light modulator (SLM). At this plane, the atmospheric turbulence was simulated for weak  ($C^2_n=1\times10^{-14}$ m$^{-2/3}$) and strong ($C^2_n=5\times10^{-14}$ m$^{-2/3}$) 
turbulence with varying link distances of 1 km, 5 km, 7 km and 20 km. Lenses, $L_3$ and $L_4$ with focal lengths of 100 mm each, were used to focus the photons into the fibre couplers ($FC_1$ and $FC_2$) which were placed on translation stages for the scanning process. The single photon pairs were detected by the coincidence counter (CC) which had embedded two single photon avalanche detector modules, $D_1$ and $D_2$.}
\label{fig:exp}
\end{figure*}

\section{Experimental Setup}
Atmospheric turbulence is a major challenge in free-space quantum communication. Mitigating these influences are necessary especially for long-distance quantum links specifically for encoding in the spatial regime. We exploit the effects of the spatial distribution of the Gaussian pump beam to generate photon-pairs with a well-defined degree of entanglement by manipulating the transverse coherence length of the pump \cite{HugoDefienne2019}. Fig. \ref{fig:exp} illustrates our experimental setup to demonstrate the influence of atmospheric turbulence on the coincidence counts between signal-idler photon pairs. A concatenated $\beta$-barium borate (BBO) 
was used to produce entangled photon pairs through a type-I SPDC process. The dotted region of the experimental set-up was used to produce the partially spatially coherence of the pump beam. A diode laser lasing at a wavelength of 405 nm was used to illuminate a rotating ground glass diffuser (RGGD) by passing through lens $L_1$ with a focal length 50 mm. The pump beam was made partially spatially coherent at the BBO crystal by placing RGGD and BBO crystal at the front and back focal plane of the lens $L_2$ ($f=40${\thinspace}mm) respectively. Non-collinear ($\pm 3\degree$ from the direction of the pump) degenerate entangled photon pairs of wavelength 810 nm were produced by pumping the BBO crystal with a partially spatially coherent pump beam. The generated photon pairs were passed through a Holoeye $1920 {\times} 1080$ spatial light modulator (SLM), which is a liquid crystal device, utilised to mimic the atmospheric turbulence strength. The signal and idler photons were detected by single photon detectors $D_1$ and $D_2$ through fibre couplers $FC_1$ and $FC_2$ respectively. The transverse spatial distribution of the photon pairs can be observed by scanning the signal-idler detector in the transverse direction. In our experiment, the coincidence counts were recorded by fixing the signal detector at $x_1=0$ and scanning the idler detector in the transverse direction $x_2$.

\section{Results and discussion}
\begin{figure*}
	\centering
	\includegraphics[width=0.8\textwidth]{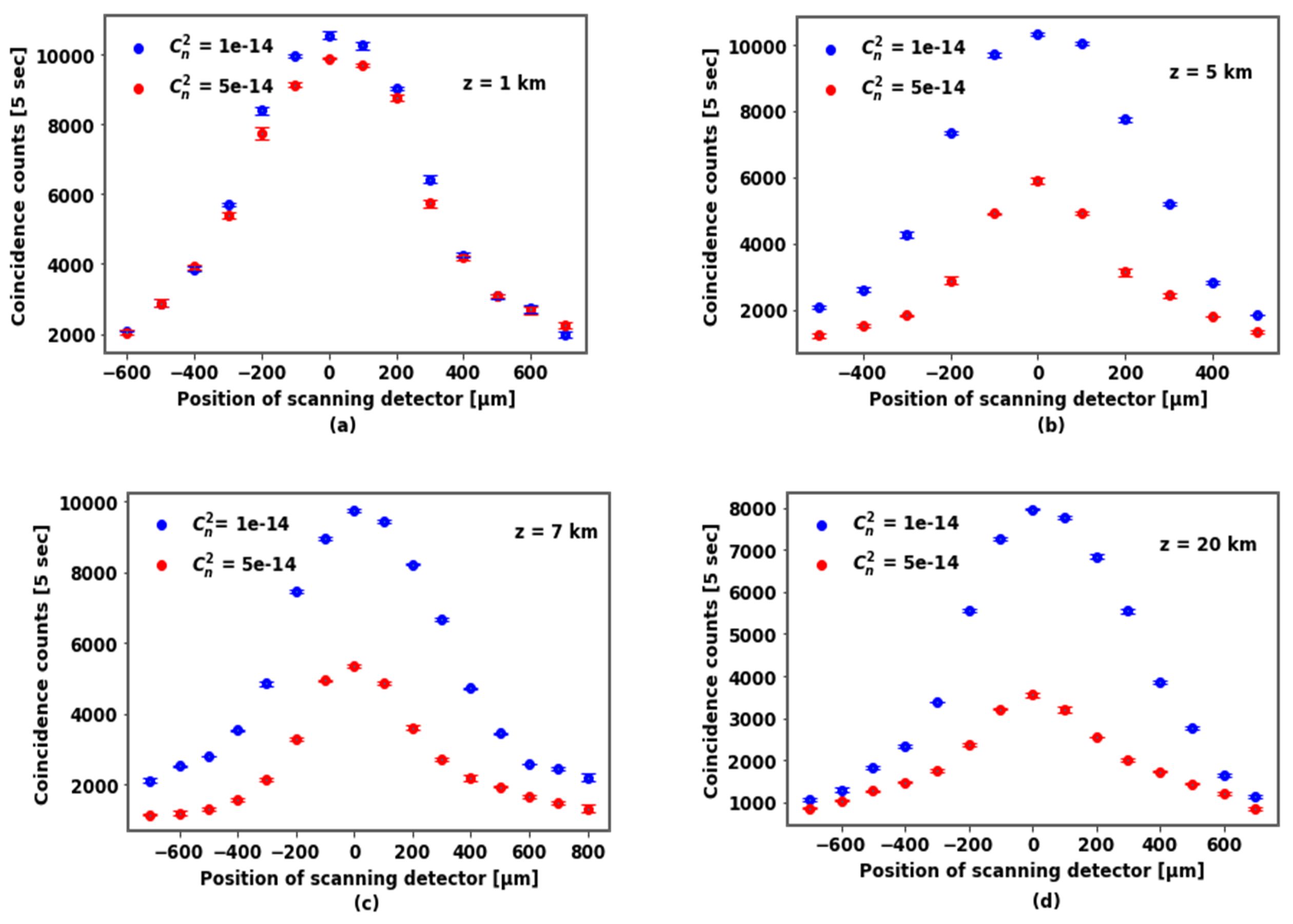}
	\caption{Experimental plots of the effect of atmospheric turbulence on the coincidence counts when pump is fully spatially coherent ($\delta=\infty$) beam for different propagation distances. (a) z=1 km, (b) z=5 km, (c) z=7 km and (d) z=20 km.}
	\label{fig:expfully}
\end{figure*}

\begin{figure*}
	\centering
	\includegraphics[width=0.8\textwidth]{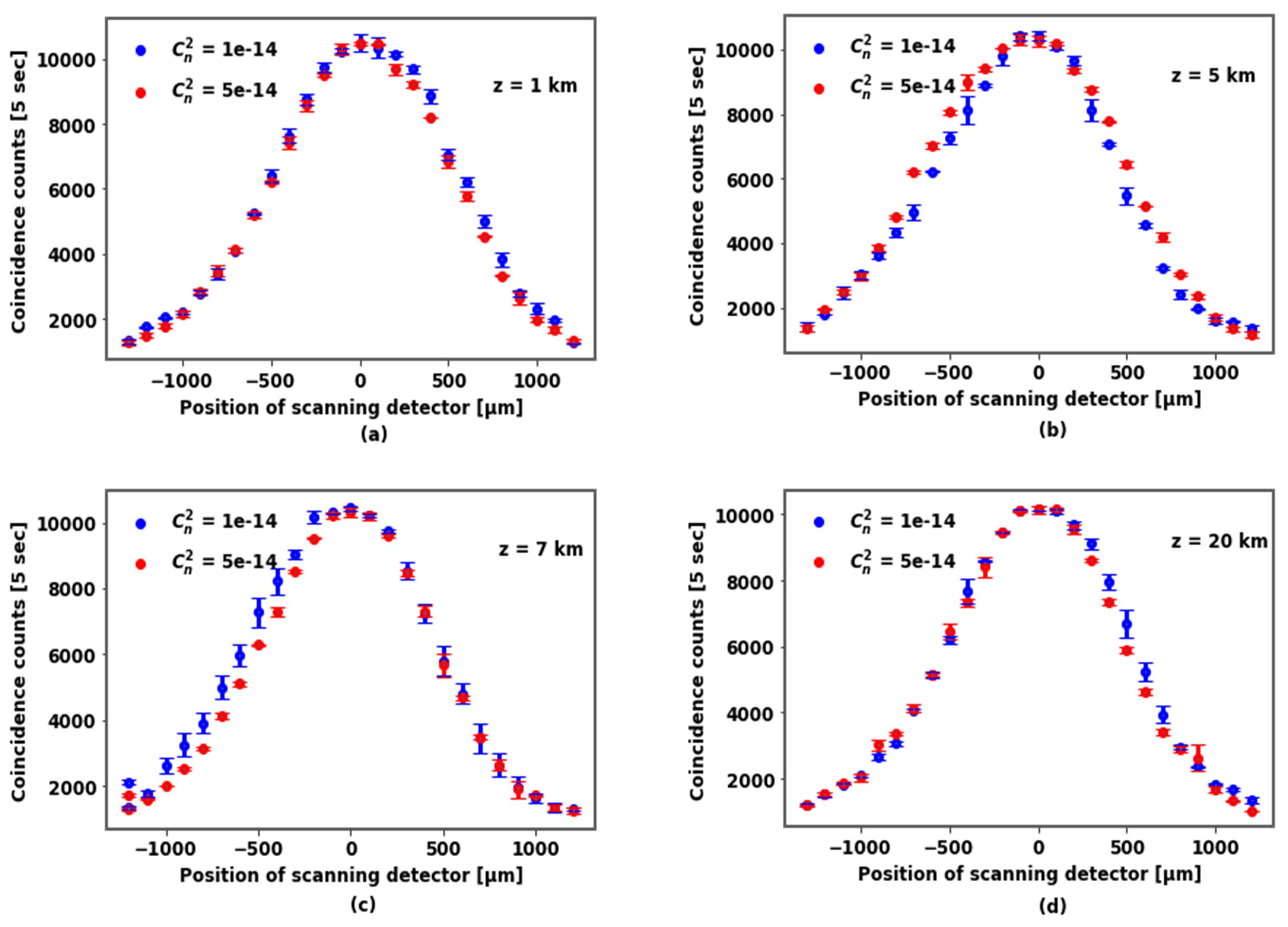}
	\caption{Experimental plots of the effect of atmospheric turbulence on the coincidence counts when pump is partially spatially coherent ($\delta=0.0876{\thinspace}$mm) for varying propagation distances. (a) z=1 km, (b) z=5 km, (c) z=7 km and (d) z=20 km.}
	\label{fig:exppartial}
\end{figure*}

\begin{figure*}
	\centering
	\includegraphics[width=0.8\textwidth]{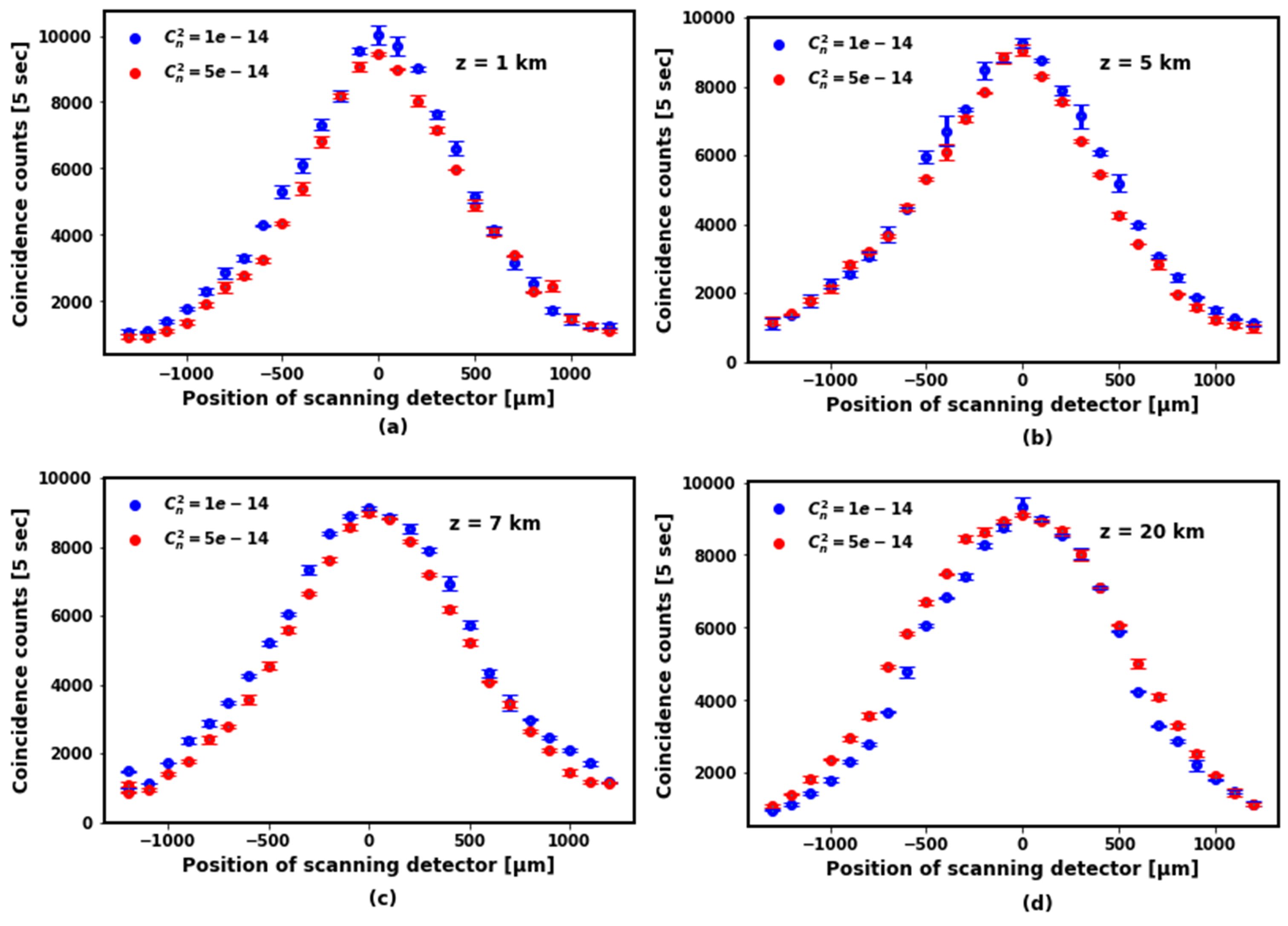}
	\caption{Experimental plots of the effect of atmospheric turbulence on the coincidence counts when pump is partially spatially coherent ($\delta=0.0417{\thinspace}$mm) for varying propagation distances. (a) z=1 km, (b) z=5 km, (c) z=7 km and (d) z=20 km.}
	\label{fig:exppartial40mm}
\end{figure*}
\begin{figure*}
	\centering
	\includegraphics[width=0.8\textwidth]{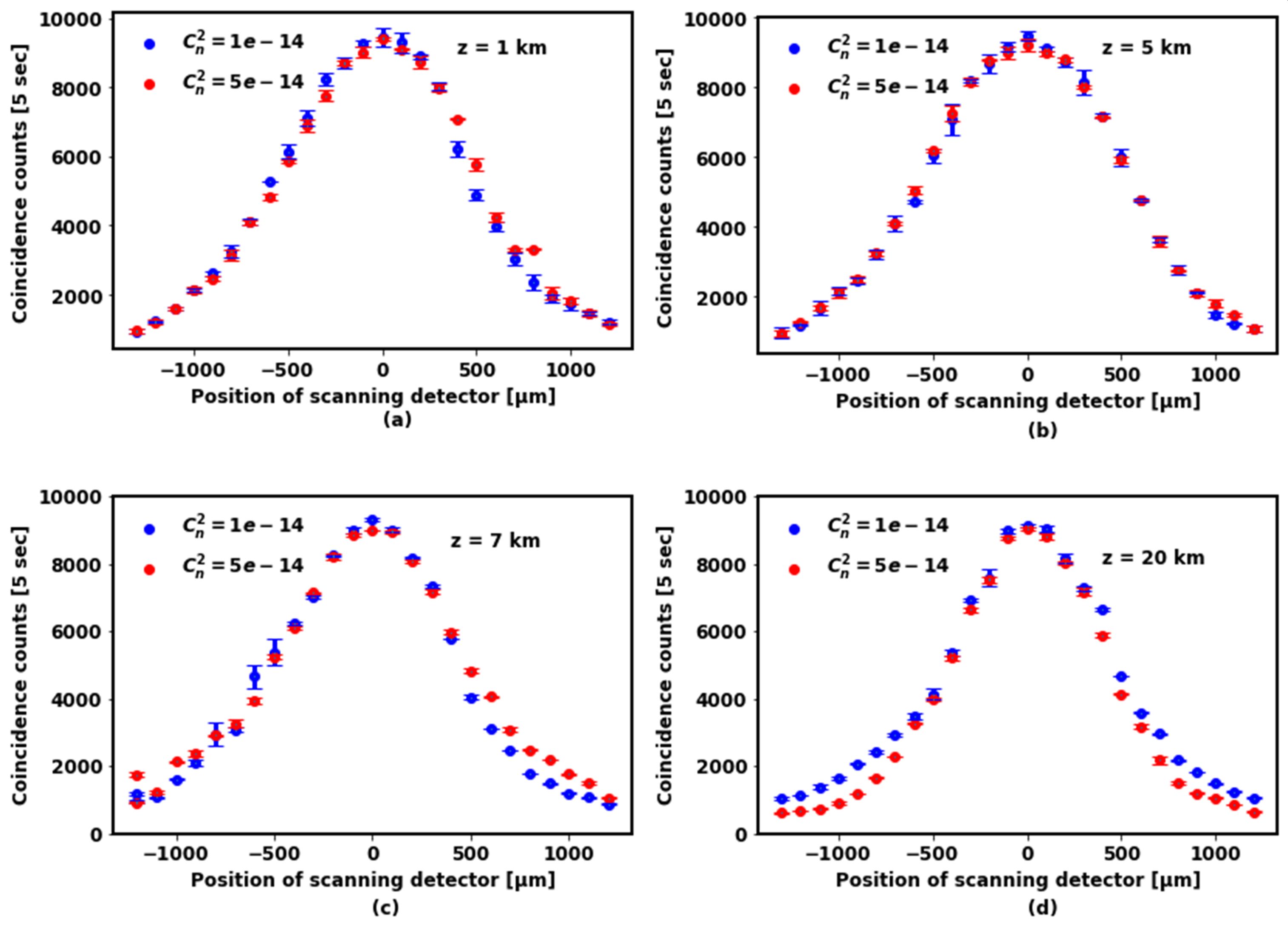}
	\caption{Experimental plots of the effect of atmospheric turbulence on the coincidence counts when pump is partially spatially coherent ($\delta=0.0253{\thinspace}$mm) for varying propagation distances. (a) z=1 km, (b) z=5 km, (c) z=7 km and (d) z=20 km.}
	\label{fig:exppartial30mm}
\end{figure*}
\begin{figure}
\includegraphics[width=0.4\textwidth]{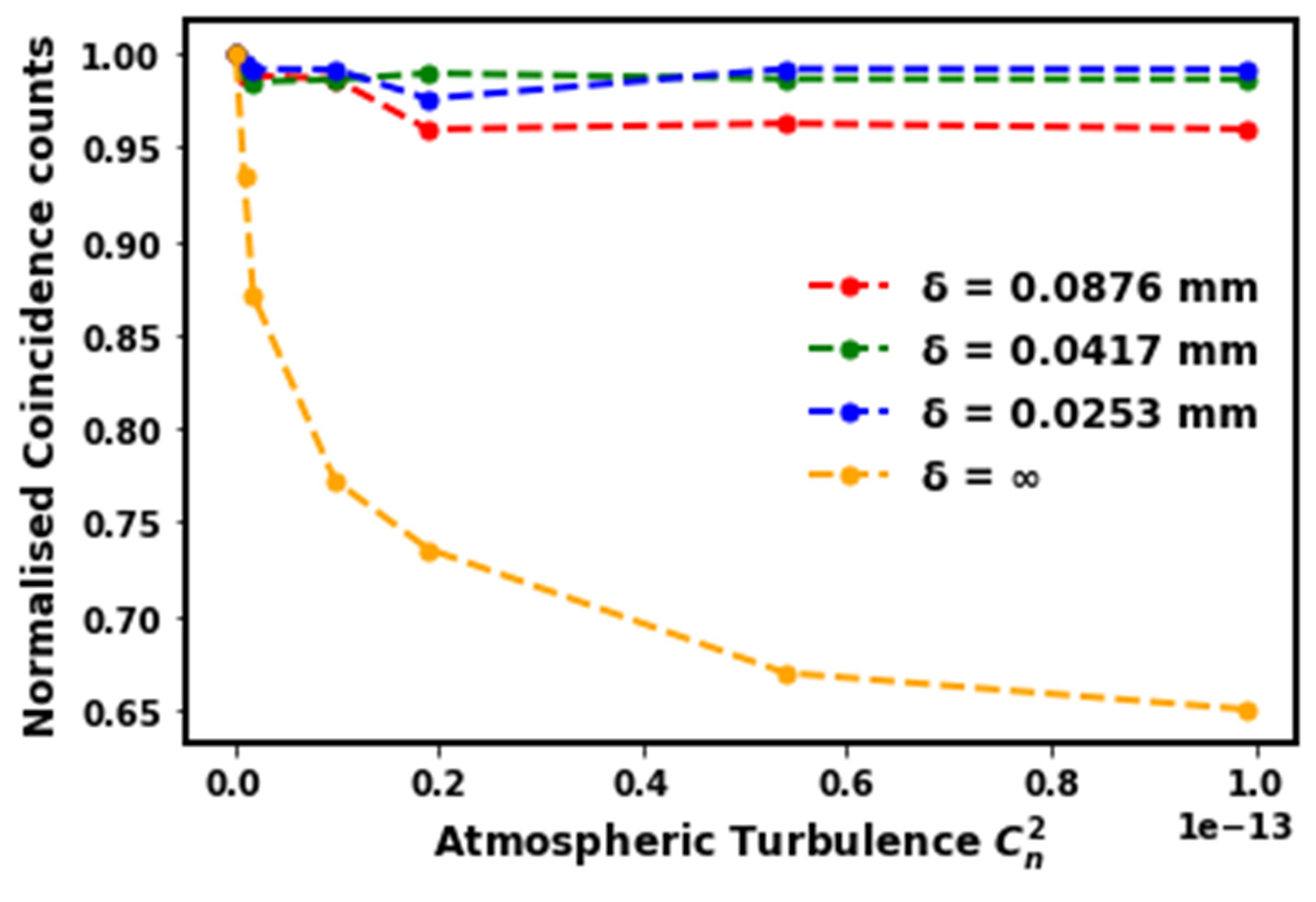}
\caption{Experimental plots of normalized coincidence counts (at $x=y=0$) as a function of $C_n^2$ at $z=20{\thinspace}$km for fully and partially coherent pump source ($\delta$ = 0.0876 mm, 0.0417 mm, 0.0253 mm).}
\label{fig:expvaryingturb}
\end{figure}

In order to study the influence of atmospheric turbulence on the signal-idler photon pairs generated by the modulated pump beam, the strength of atmospheric turbulence was varied from $C^2_n=10^{-14}$ m$^{-2/3}$ (weak turbulence) to $C^2_n=5{\times}10^{-14}$m$^{-2/3}$ (strong turbulence). 
The theoretical plots in Fig. \ref{fig:theoryfully} and Figs \ref{fig:theorypartial} - \ref{fig:theorypartial3} show the influence of atmospheric turbulence on the coincidence counts as a function of scanning detector position $x_2$ for fully ($\delta=\infty$) and partially ($\delta=0.0876${\thinspace}mm,{\thinspace} $0.0417${\thinspace}mm and $0.0253${\thinspace}mm) spatially coherent pump beam respectively. 
To demonstrate the influence of atmospheric turbulence on the coincidence detection experimentally, the SLM was encoded for weak and strong atmospheric turbulence strength based on Kolmogorov model \cite{kolmogorov21941,kolmogorov1941}. Furthermore, the quantum link was varied in a number of distances ranging from 1-20km.
\newline
\indent
Initially, the fully spatially coherent pump beam was used to illuminate the BBO crystal in the absence of the diffuser system. 
The generated signal-idler photon pairs were passed through the SLM and detected by fixing the signal at $x_1=0$ and scanning the idler in the transverse direction $x_2$. Fig. \ref{fig:expfully} shows the experimental results depicting the influence of the atmospheric turbulence on the coincidence counts at different propagation distances. 
Fig. \ref{fig:theoryfully}a and Fig. \ref{fig:expfully}a  show a small difference in coincidence detection for the two ranges of $C^2_n$. This implies that when the propagation distance is less than 1 km, the effect of turbulence on the coincidence detection is small. As the photons propagate over longer distances, the difference in coincidence detection increases for the two values of $C^2_n$ (strong and weak), as shown in Fig. \ref{fig:theoryfully}b - Fig.    \ref{fig:theoryfully}d and Fig. \ref{fig:expfully}b - Fig. \ref{fig:expfully}d.
The variation in maximum coincidence count detection for a fully coherent pump source at 1 km was 10 $\%$ (Fig. \ref{fig:expfully}a) between strong and weak turbulence. This variation reached 56 $\%$ at 20 km (Fig. \ref{fig:expfully}d). Therefore, the detection of spatially generated photons becomes more difficult as they propagate through a longer distance in stronger turbulence.
\newline
\indent
For the variation of the partial coherence, the pump spatial coherence length ($\delta$ = 0.0876{\thinspace}mm, 0.0417{\thinspace}mm, 0.0253{\thinspace}mm) was varied  by increasing the spot size at the RGGD as shown in the dotted part in the experimental system in Fig. \ref{fig:exp}. The coherence length at the plane of BBO crystal was calculated by, $\delta=\frac{3.832{\lambda}f}{2{\pi}d}$ \cite{foley1991effect}, where ${\lambda}$ is the wavelength of the pump beam, $f$ is the focal length of lens $L_2$ and $d$ is the spot size at the diffuser plane. The theoretical plots in Figs. \ref{fig:theorypartial} - \ref{fig:theorypartial3} and the corresponding experimental results in Figs. \ref{fig:exppartial} - \ref{fig:exppartial30mm} illustrate the influence of atmospheric turbulence on the coincidence detection, when the signal-idler photon pairs were produced by a partially spatially coherent pump ($\delta=0.0876${\thinspace}mm, 0.0471{\thinspace}mm, 0.0253{\thinspace}mm respectively), for different propagation distances. It is observed that the same coincidence detection was acquired for the two ranges of $C^2_n$ (strong and weak atmospheric turbulence) for short and longer propagation distances.
This implies that the photon pairs produced with a low spatial coherence pump are less affected by the atmospheric turbulence strength. It was observed that for all of the aforementioned distances, the signal-idler photon pairs generated by the partially spatially coherent pump are less susceptible to atmospheric turbulence than the photon pairs produced with fully spatially coherent pump.
\newline
\indent
To emphasise the advantage of the partial spatial coherence for varying atmospheric turbulence, the experimental observation of the normalized coincidence counts detected is further summarized in Fig. \ref{fig:expvaryingturb} for a propagation distance of 20 km. It was observed that for the fully coherent pump source (yellow dashed curve) there is an exponential decay in coincidence count detection as the atmospheric turbulence strength is increased. For the partially coherent pump source ($\delta$ = 0.0876 mm, 0.0417 mm, 0.0253 mm) the coincidence count detection is almost constant (red, blue and green dashed curve respectively). By modifying the spatial coherence of the pump beam, we produce entangled photons which are considered weak measurements however even though this has implications on the strength of the entanglement \cite{HugoDefienne2019} it does prove to be resistance to varying atmospheric turbulence. For a free-space quantum communication link, this can be circumvented by increasing the transmission time and hence sending more photons through the quantum channel to produce the secure key.
\section{Conclusion}
In conclusion, we have theoretically and experimentally demonstrated the influence of atmospheric turbulence on the coincidence detection of entangled photons produced by a fully and partially coherent pump source. To the best of our knowledge, this is the first observation of the propagation of entangled photons through atmospheric turbulence using partially spatially coherent pump source. To simulate the free-space quantum link, a spatial light modulator encoded with the Kolmogorov model was used to mimic the atmospheric conditions. The RGGD was used in combination with a lens in a $2f$ geometry to produce the partially spatially coherent pump beam at the BBO crystal plane. It was found that the coincidence counts remained the same for weak and strong turbulence when the signal-idler photons pairs were generated using a partially spatially coherent pump. However, there is a decrease in coincidence counts for strong turbulence when entangled photons are produced with a fully spatially coherent pump beam. This implies that the entangled-photon pairs generated using a partially spatially coherent pump are more robust towards varying atmospheric turbulence strength than the entangled-photons generated using a fully spatially coherent pump. Despite it being suggested that the partial spatial coherent pump have implications on the single photon generation, however, there is a significant resilience toward atmospheric turbulence, especially for longer propagation distances. These result may find important applications in free-space quantum communication using spatially entangled photons.

\section*{Acknowledgement}

This work is based on research supported by the South African Research Chair Initiative of Department of Science and Technology and National Research Foundation as well as the Thuthuka grant. Opinions expressed and conclusions arrived at, are those of the authors and are not necessarily to be attributed to the NRF. S J acknowledges Indian Institute of Technology Delhi India for financial support as post-doctoral fellowship.
\bibliographystyle{apsrev4-1}
\bibliography{manuscript.bib}
\end{document}